\documentclass[11pt]{article}
\usepackage{moriond,epsfig}

\def\be{\begin{equation}}
\def\ee{\end{equation}}
\def\bea{\begin{eqnarray}}
\def\eea{\end{eqnarray}}

\begin{document}
\vspace*{2.cm}
\title{BEAUTY PRODUCTION CROSS SECTION \\
MEASUREMENTS AT E$^{}_{cm}$ = 1.96 TeV}

\author{ MONICA D'ONOFRIO \\
(On behalf of the CDF and D0 collaborations)}

\address{D\'epartement de physique nucl\'eaire et corpusculaire, \\
University of Geneva, Quai Ansermet 24, 1205 Geneva, Switzerland}

\maketitle\abstracts{
The RunII physics program at the Tevatron started in spring 2001 with protons and antiprotons colliding 
at an energy of $\sqrt{s}$=1.96 TeV, and it is carrying on with more than 500 pb$^{-1}$ of data as collected 
by both the CDF and D\O \, experiments. Recent results on beauty production cross section measurements are here reported.} 
\section{Introduction}
Beauty production measurements provide precision tests of perturbative QCD. At the Tevatron, the total $b\bar{b}$ cross section 
is about 50 $\mu$b, resulting in an event rate of few kHz, thus, in a very high statistics.  
In the past years, the publication of "controversial" beauty cross section measurements by CDF and D\O \, in RunI, 
has led to many developments both in the theoretical calculations beyond NLO and the experimental approach, 
resulting in a better agreement between data and theory as it is shown in this paper.   
A major role, concerning theoretical improvements, has been played by the implementation of the so called Fixed-Order with 
Next-to-Leading-Log (FONLL) calculation~\cite{mangano}. 
 Also, there have been substantial changes in bottom fragmantation function as extracted from experimental data and for the used 
Parton Distribution Functions (CTEQ6M). 
On the other hand, from the experimental side, there have been many improvements in the treatment of data inputs, 
for instance avoiding deconvolution and extrapolation to parton level with Monte Carlo simulations, 
and implementing the use of real observables as b-hadrons and b-jets.    
\section{B hadron production cross section using J/$\psi$}
\begin{figure}[h!]
\mbox{
\begin{minipage}{0.45\textwidth}
\centerline{\mbox{\psfig{figure=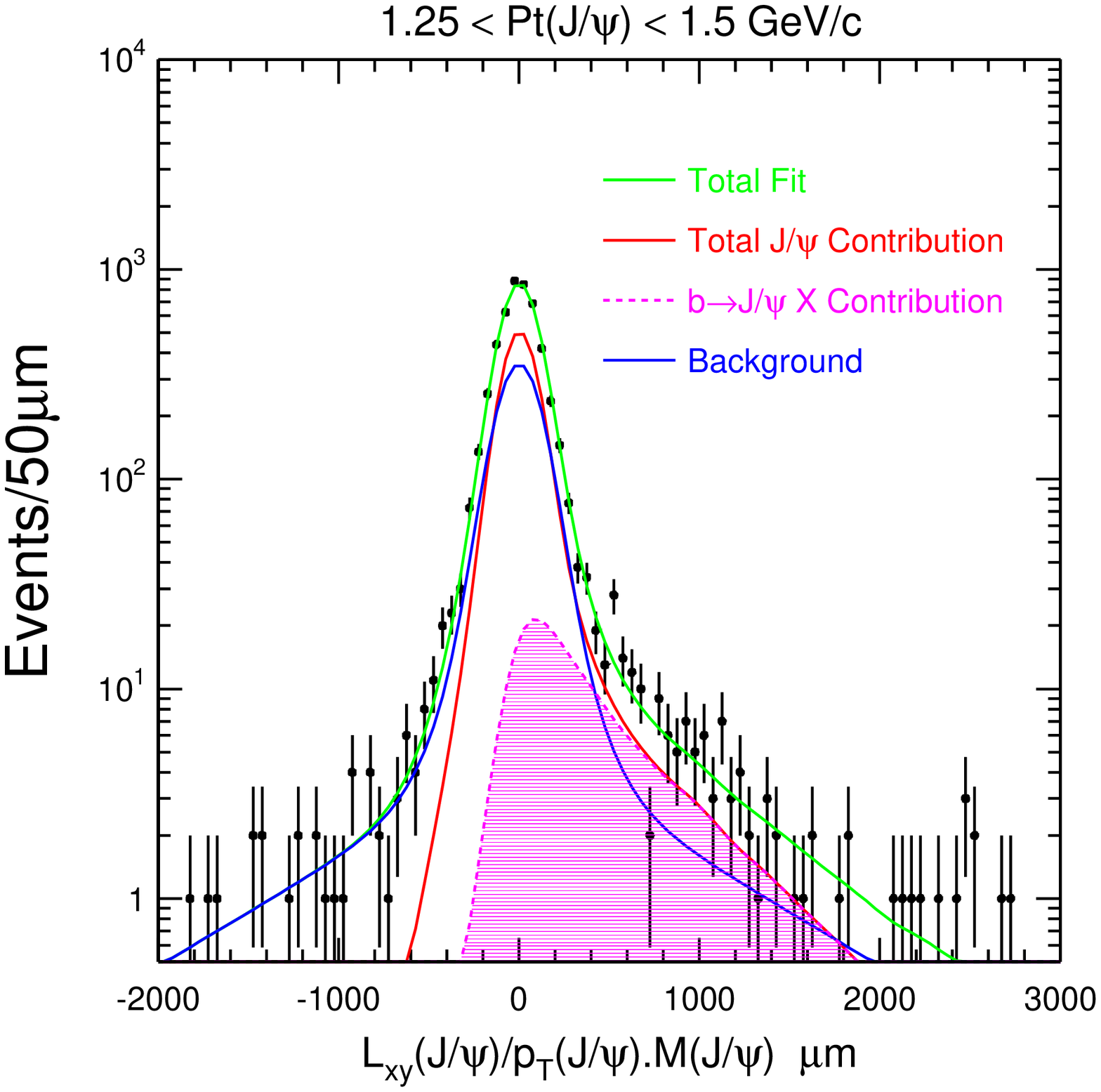,height=1.8in,width=1.\textwidth}}}
\vspace{-0.05in}
\caption{Example of fits to the J/$\psi$ pseudo-proper decay time in the range 1.25$<p^{\mu \mu}_{T}<$1.5 GeV/$c$ 
to extract J/$\psi$ from long-lived B-hadron decays. 
\label{fig:bhadr_mass}}
\end{minipage}\hspace*{0.4in}
\begin{minipage}{0.45\textwidth}
\centerline{\mbox{\psfig{figure=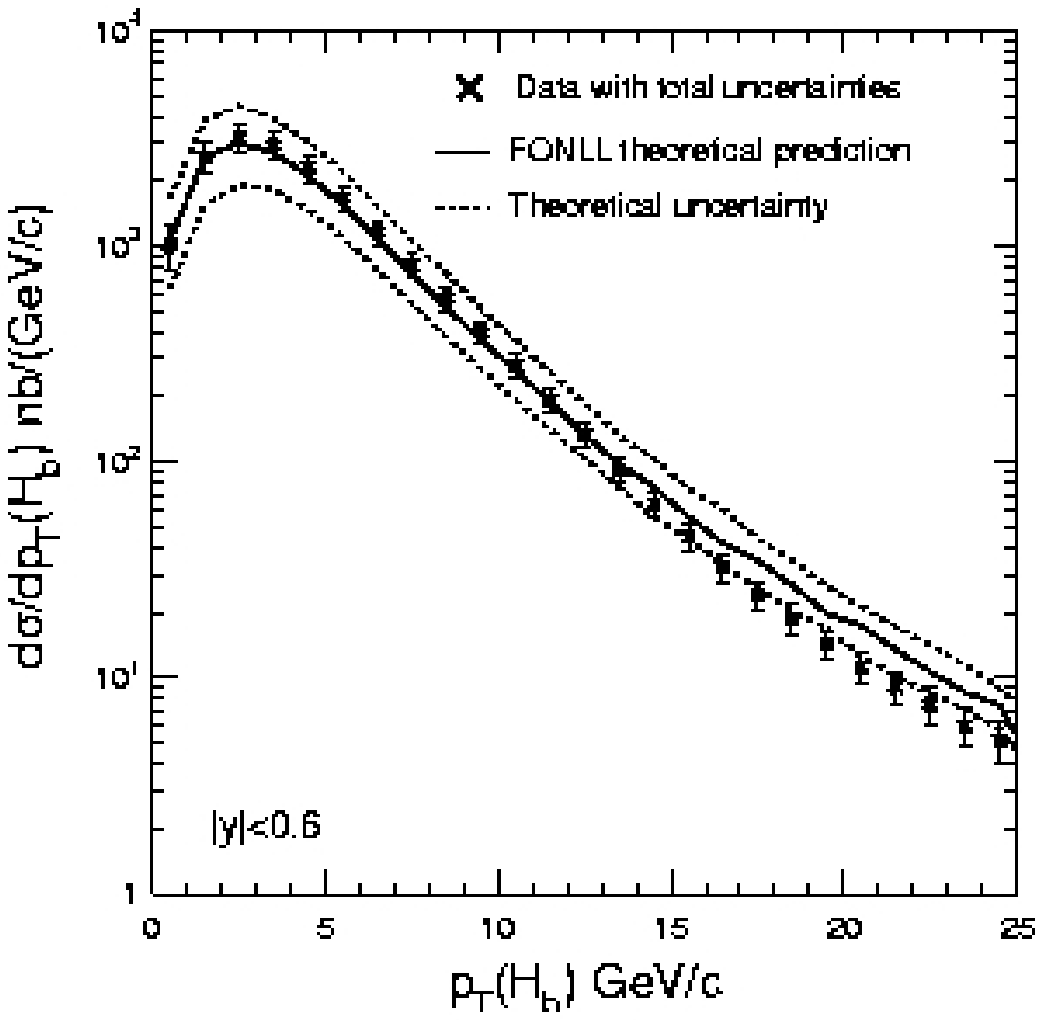,height=1.6in,width=1.\textwidth}}}
\vspace{-0.05in}
 \caption{B-hadron production cross section as a function of 
B-hadron p$^{}_{T}$. The crosses with erros bars are the data 
with systematic and statistical error added. Theoretical prediction is superimposed.}
\label{fig:bhadr_cross}
\end{minipage}
}
\end{figure}

A measurement of the B-hadron ($H^{}_{b}$) production cross section~\cite{mary} 
has been performed by CDF, using data corresponding to an integrated luminosity 
of 39.7 pb$^{-1}$. First, the inclusive production cross section of J/$\psi$ is measured, 
considering J/$\psi \, \rightarrow \, \mu^{+} \mu^{-} $. This   
cross section contains contributions from various sources, including prompt 
production of charmonium, decays of excited charmonium states and decays 
of B-hadrons; once the latter J/$\psi$ component is extracted, it is possible 
to determine in a very clean way the B-hadron cross section. \\ 
A di-muon trigger is used to select the events: two oppositely charged muons 
are required to be reconstructed from tracks measured in the tracking 
system and matched to track segments detected in the muon chambers. 
Central muons ($|y^{\mu}_{}|<$0.6) with p$^{}_{T}>$1.5 GeV/$c$ are used: the offline reconstruction efficiency is about 99$\%$ 
indipendently by the p$^{\mu }_{T}$, thus the measurement of J/$\psi$ cross section covers all transverse momenta from 0 to 20 GeV/$c$ 
in the central rapidity range $|y^{J/ \psi}_{}|<$0.6.              
To separate prompt J/$\psi$ and J/$\psi$ from decay 
$H^{}_{b} \, \rightarrow J/\psi X$, lifetime distributions are used. 
B-hadrons have long lifetimes, on the order of picoseconds, so that 
J/$\psi$ from those decays are usually displaced from the primary vertex, 
taken as the beam position in the $r-\phi$ plane. 
The signed projection of the flight distance of J/$\psi$ on its transverse momentum, 
L$^{}_{xy}$, is, in general, a good measurement of the displaced vertex; nevertheless, to reduce 
the dependence on the J/$\psi$ p$^{}_{T}$ bin, the variable 
$x = L^{}_{xy} \times (M^{J/\psi}_{}/p^{J/\psi}_{T}) $, called pseudo-proper 
decay time, is used. 
A Monte Carlo simulation is implemented to model the $x$ distributions from B-hadron events 
and, finally, an unbinned maximum likelihood fit is used to extract the B fraction 
from the data for different ranges in $p^{J/\psi}_{T}$. 
Figure~\ref{fig:bhadr_mass} shows an example of fits to the pseudo-proper decay time; the B fraction 
varies from 10$\%$ (low p$^{}_{T}$) to 45$\%$ (high p$^{}_{T}$). The extracted differential cross section 
of the B-hadrons over the transverse momentum range from 0 to 25 GeV/$c$ is shown in figure~\ref{fig:bhadr_cross}. 
The cross section is corrected for the branching fraction $Br(H^{}_{b} \rightarrow J/\psi X)$ = 1.16$\pm$0.10$\%$ and $Br(J/\psi \rightarrow \mu^{+}\mu^{-})$
 = 5.88$\pm$0.10$\%$, and divided by two to have the single B-hadron cross section. The most recent theoretical calculations are also superimposed, 
showing a remarkable agreement with data. 
\section{B-jet production}
\begin{figure}[t]
\centerline{\epsfig{file=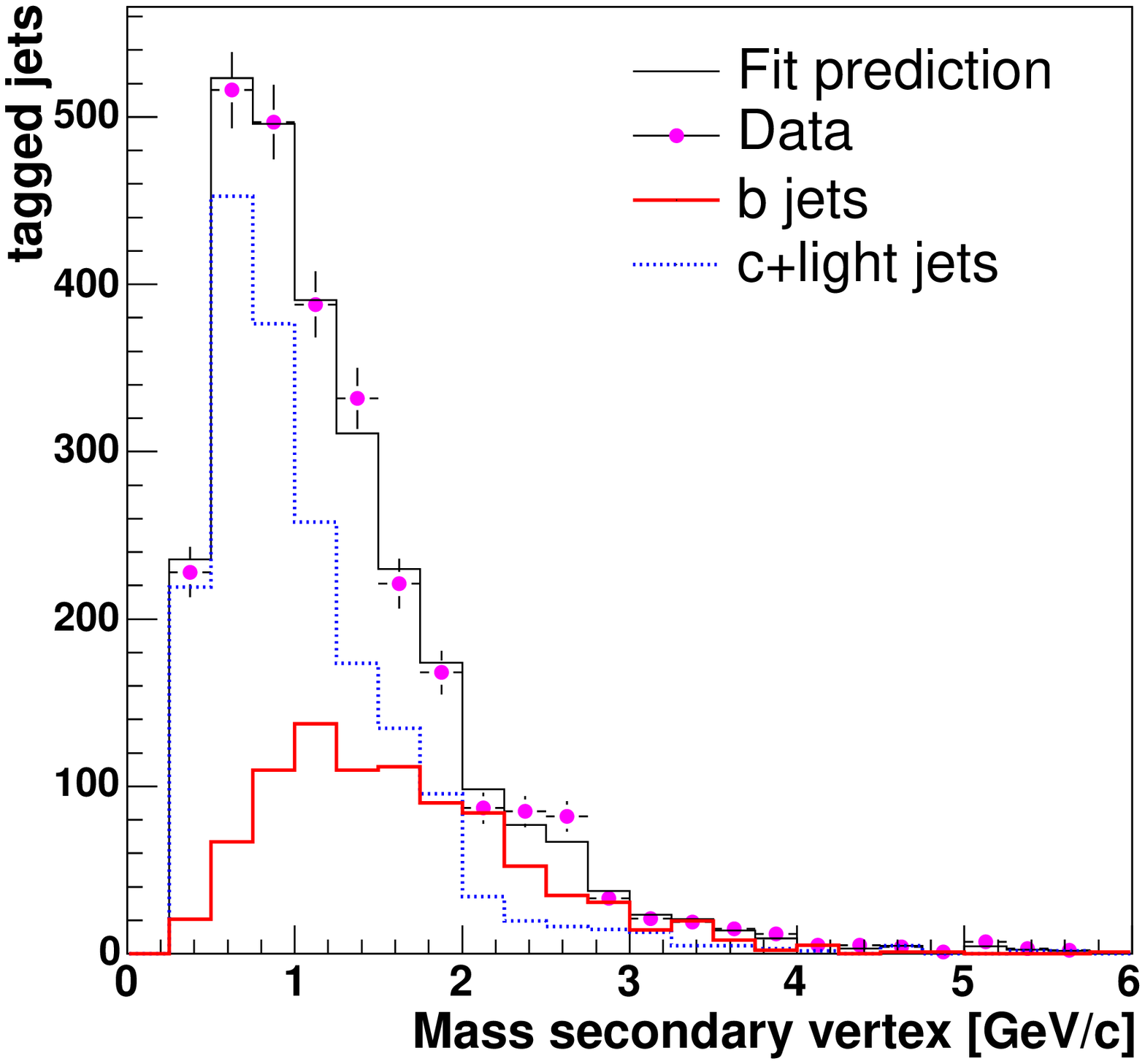,height=1.4in,width=0.45\textwidth}
\epsfig{file=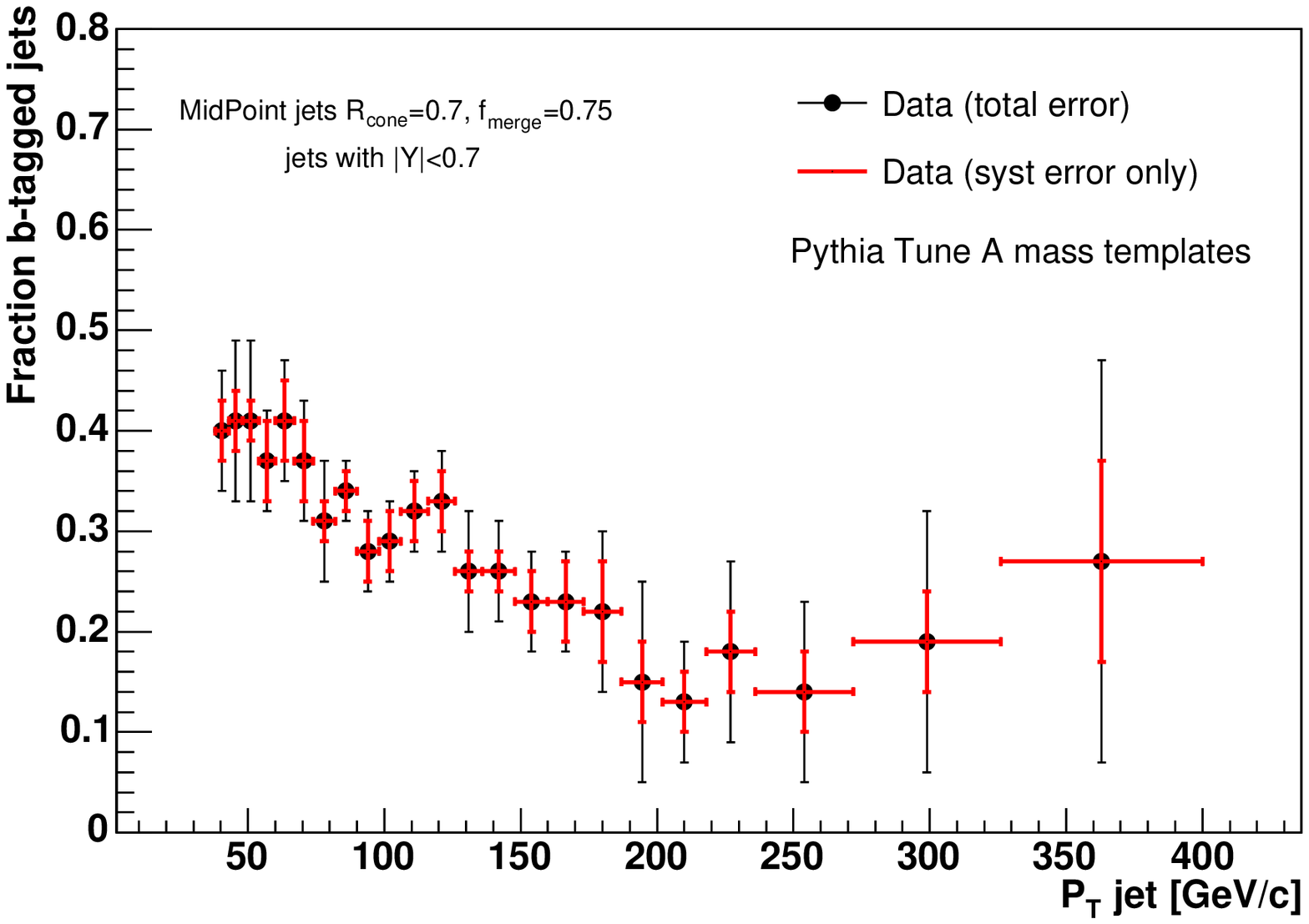,height=1.4in,width=0.45\textwidth}}
\caption{\emph{Left}: Example of mass of secondary vertex distributions for data and fit prediction 
superimposed to Monte Carlo predictions. for one bin in jet P$^{}_{T}$ 
(82$<p^{jet}_{T}<$90 GeV/$c$). \emph{Right}: Fraction of b-tagged jets as a function of corrected jet P$^{}_{T}$: total errors and systematic uncertainties are superimposed. \label{fig:bjet_mass}}
\end{figure}

The measurement of the inclusive $b$-jet cross section~\cite{me} in RunII (CDF),     
is based on about 300 pb$^{-1}$ of data. The use of jets extends the upper reach 
of the beauty production measurement performed using exclusive 
decays (P$^{Bhadr}_{T}<$ 25 GeV/c). Besides, theoretical uncertainties 
on fragmentation are rather small.
A cone-based iterative MidPoint algorithm is used for 
jet reconstruction in the Y-$\phi$ space, using a cone radius of 0.7. Only jets in the central 
rapidity region ($|$y$|<$0.7) are considered. The jet energy scale is corrected, using a Monte Carlo simulation,  
to compansate for energy losses at calorimeter level. \\ 
The analysis exploits the good tracking capabilities of the detector and rely on b-jet identification  
done by secondary vertex reconstruction. 
The b-tagging algorithm uses displaced tracks associated 
with a jet that are within the jet cone. The search for secondary vertices is defined in two steps,  
with selection based on the significance of the impact parameter and of
the decay lenght L$^{}_{xy}$. The sign of L$^{}_{xy}$ is used as further selection to reject mistagged jets.  
To determine the heavy flavour content of a tagged jet, thus to 
extract the fraction of b-jets, the shape of the secondary vertex mass distribution 
is used: altough a full reconstruction of the hadron invariant mass 
is not possible, due to the presence of neutral particles and energy lost because of 
detector resolution, the invariant mass of tracks used to find the 
secondary vertex constitutes a good discrimination for jets from long-lived b or c hadrons. 
The mass of secondary vertex depends on jet transverse momentum and the fit is performed 
considering independently each bin in jet P$^{}_{T}$: in figure ~\ref{fig:bjet_mass}, 
on the left, mass distribution and fit results are presented for one bin in jet P$^{}_{T}$ as example.  
On the right, the fraction of b-tagged jets as a function of jet P$^{}_{T}$ is showen: 
the total errors and systematic uncertainties are superimposed.
\begin{figure}[h]
\centerline{\epsfig{file=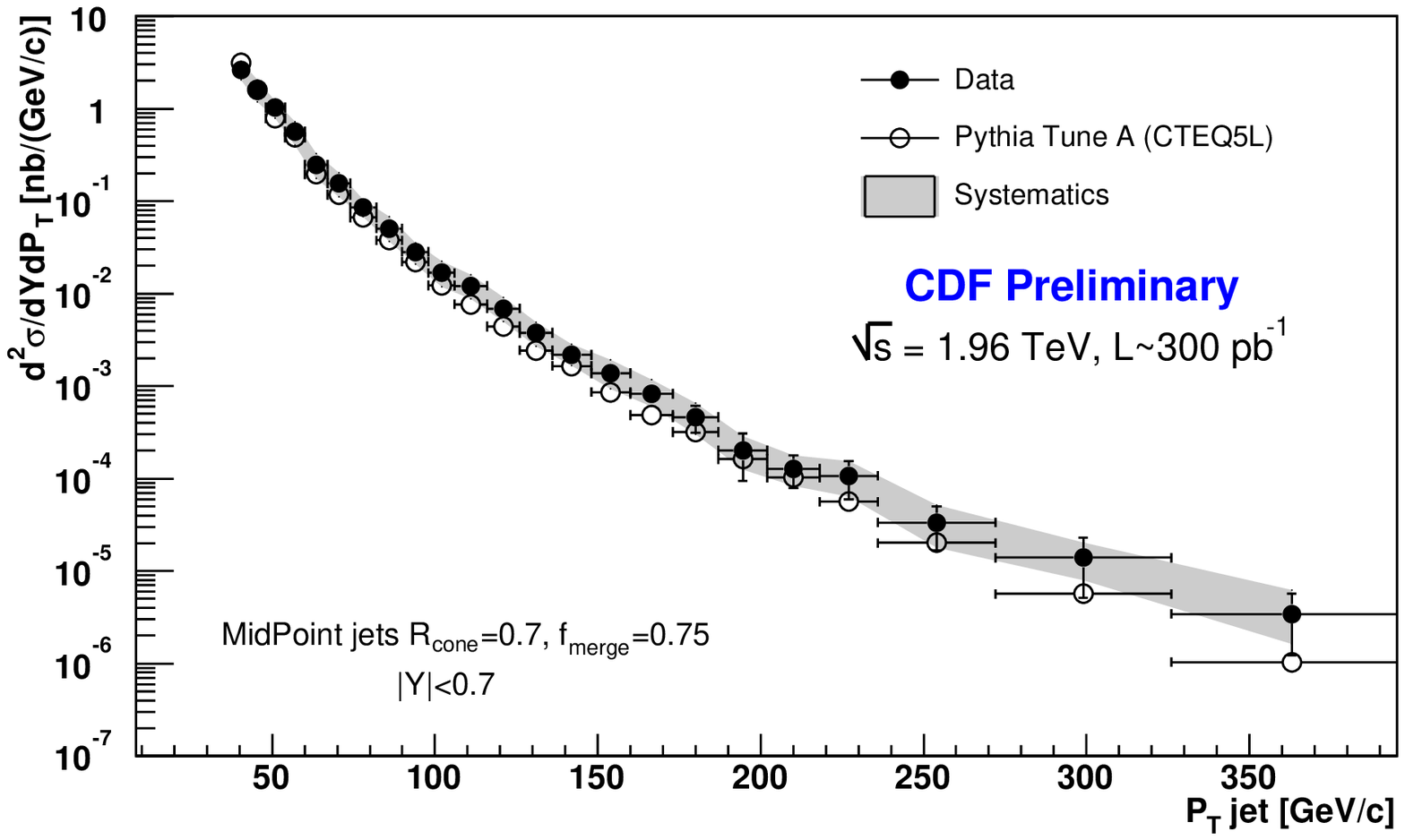,height=1.9in,width=0.49\textwidth}
\epsfig{file=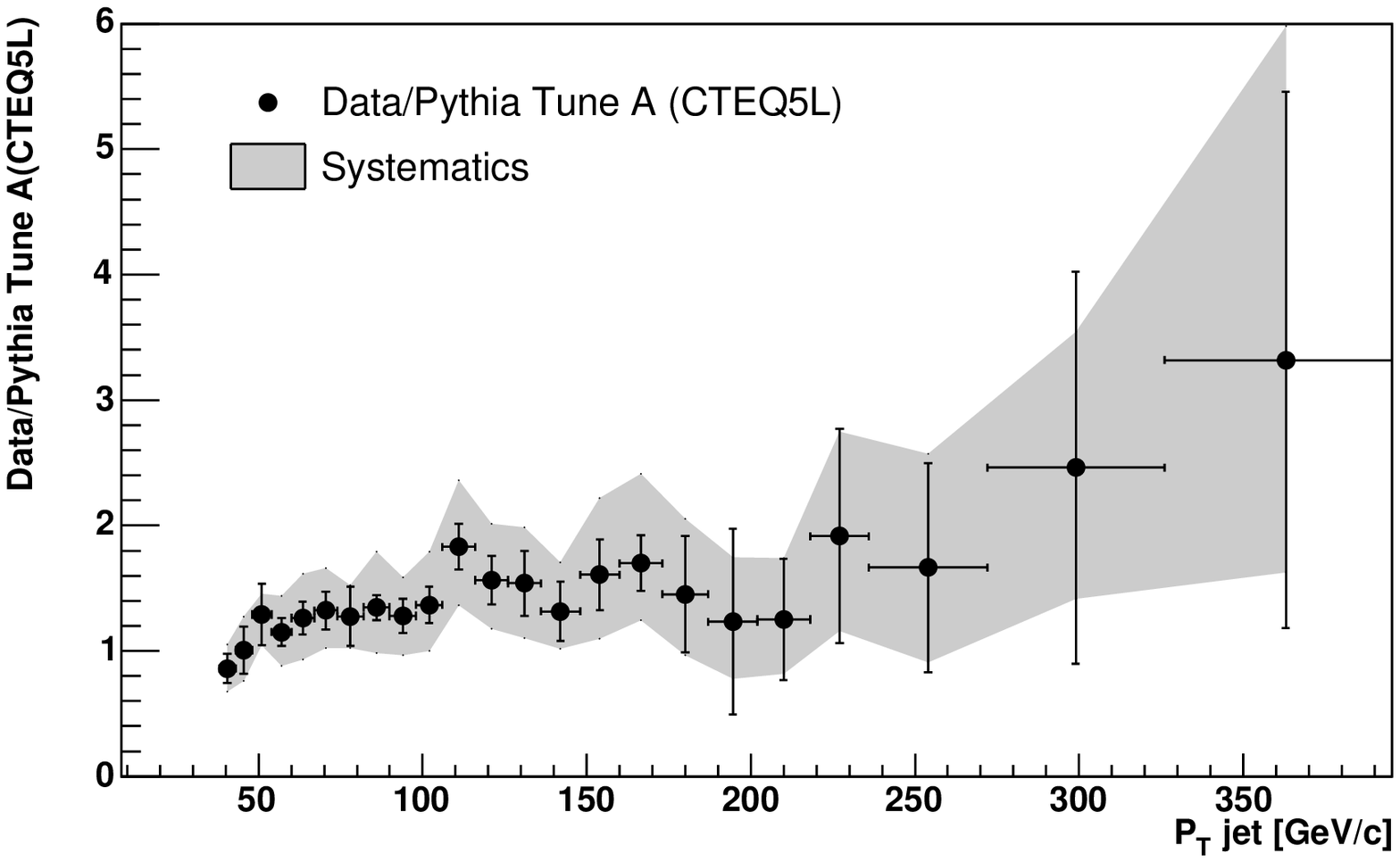,height=1.9in,width=0.49\textwidth}}
\caption{\emph{Left}:  b-jet cross section as a function of corrected jet P$^{}_{T}$, 
superimposed to Pythia Monte Carlo LO(CTEQ5L) prediction. \emph{Right}: Ratio Data/Pythia  
as a function of corrected jet P$^{}_{T}$. \label{fig:bjet}}
\end{figure}
Figure~\ref{fig:bjet} shows the inclusive b-jet cross section over a P$^{}_{T}$ range between 
38 and 400 GeV/$c$.
Statistical and systematic error on the last bins is dominated by the error on b-tagged jets fraction. One of the main systematic 
error contributions comes from the jet energy scale: a conservative 5$\%$ uncertainty is considered at this stage, although on-going studies 
show the possibility to reduce it at 3$\%$. The b-jet cross section is compared with leading order (LO) 
prevision from Pythia Monte Carlo (CTEQ5L), and results are in agreement with expectation: 
a NLO and beyond comparison is in progress. \\  
D\O \, is also performing a b-jet cross section measurement using a muon-tagging method to select jet sample     
enhanced in heavy flavor content: work is in progress to separate the b-jet content from the background.  
\section{Upsilon production}
\begin{figure}[t]
\mbox{
\begin{minipage}{0.45\textwidth}
\centerline{\mbox{\psfig{figure=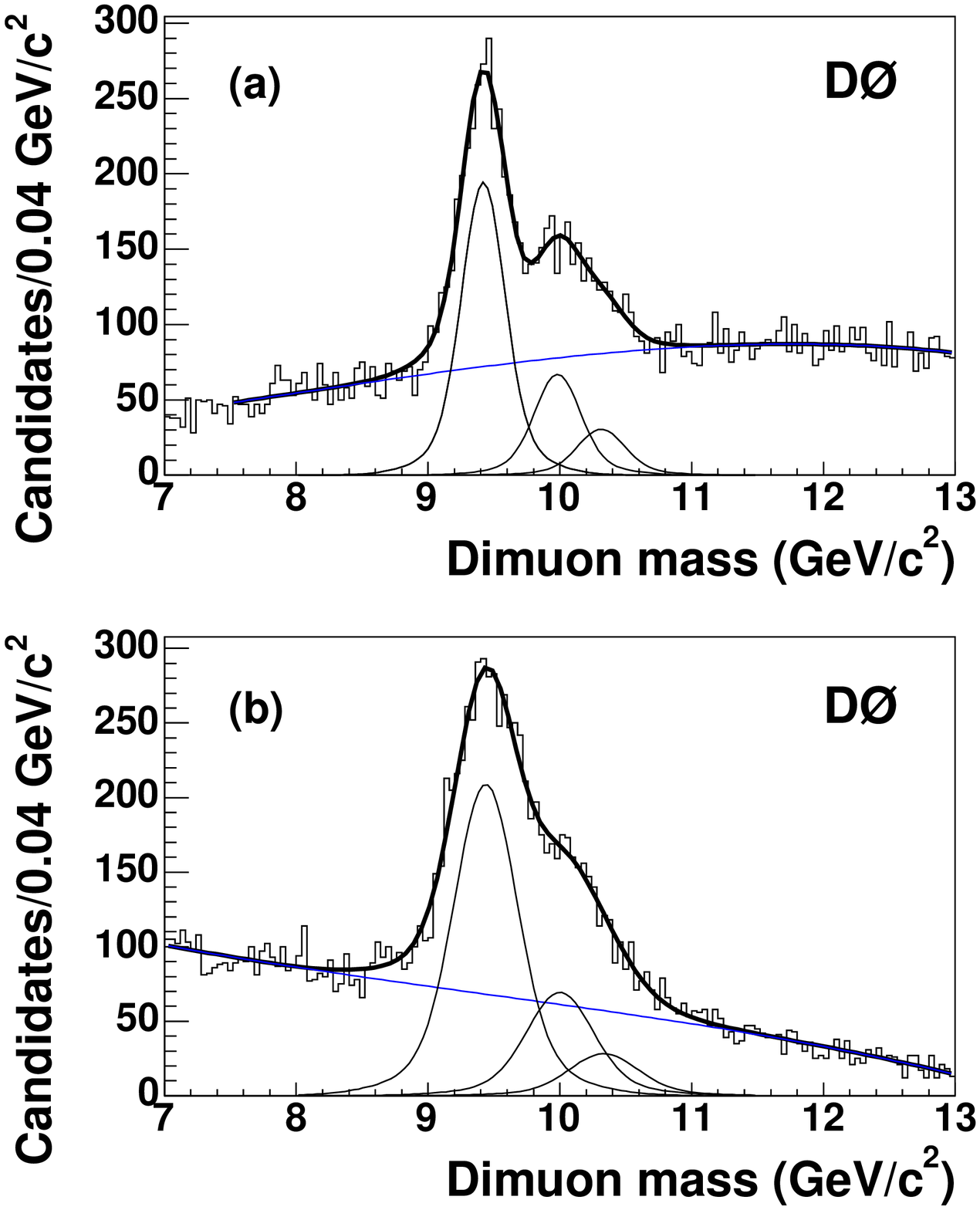,width=6.3cm,height=5.7cm}}}
\vspace{-0.05in}
\caption{Examples of di-muon mass distributions in rapidity range $|y^{\Upsilon}|<$0.6 
and 1.2$\leq |y^{\Upsilon}| \leq$1.8, for $p^{\Upsilon}_{T}$ range [4,6] GeV/$c$. 
The heavy line is the combined fit for signal and background.}
\label{fig:upsi_mass}
\end{minipage}\hspace*{0.4in}
\begin{minipage}{0.45\textwidth}
\centerline{\mbox{\psfig{figure=moriond_plots/cross_bottomon.epsi,width=6.3cm,height=6.cm}}}
\vspace{-0.05in}
 \caption{\emph{Top}: Normalized differential cross sections for $\Upsilon$(1S) 
   production. Theoretical prediction are superimposed. 
   \emph{Bottom}: Ratio of cross sections for  
   $\sigma$(1.2$\leq |y^{\Upsilon}| \leq$1.8) to $\sigma$($|y^{\Upsilon}|<$0.6). Monte Carlo prediction is superimposed.}
 \label{fig:upsi_cross}
\end{minipage}
}
\end{figure}

The D\O \, experiment has performed a precise measurement of the production  
of the $\Upsilon$(1S) bottomonium state~\cite{ups}, reconstructed through its decay 
$\Upsilon$(1S)$\rightarrow \, \mu^{+} \mu^{-}$. The differential cross section  
as a function of the $\Upsilon$ transverse momentum is determined in three rapidity 
ranges, \begin{math} 0<|y^{\Upsilon}|\leq0.6 \end{math}, 0.6$\leq |y^{\Upsilon}| \leq$1.2, and 
1.2$\leq |y^{\Upsilon}| \leq$1.8. 
The data used in this analysis correspond to 159 pb$^{-1}$, collected 
with a scintillator-based di-muon trigger; two opposite-charge 
muon candidates are required, each of them with $p^{}_{T}>3$ GeV/$c$ and 
$|y^{\mu}|<$2.2, and matched to a track found in the central tracking system.    
Isolation requirements are imposed to reduce the dominant background 
from $b\bar{b}$. 
Two typical examples of di-muon mass distribution in different 
bins of rapidity and in the $p^{}_{T}$ bin of 4 GeV/$c$$<p^{\Upsilon}_{T}<$6 GeV/$c$  
are showed in~\ref{fig:upsi_mass}: the clear $\Upsilon$(1S) signal is accompained by a 
shoulder from $\Upsilon$(2S) and $\Upsilon$(3S). 
The mass distribution fits use three separate mass functions for each of the $\Upsilon$(\emph{n}S) states. 
The overall systematic errors are about 10$\%$, excluding luminosity 
uncertainty that is 6.5$\%$: the fitting procedure is one of the main 
source of uncertainties, while others as momentum resolution, track quality and track 
matching requirements contribute for less than 2$\%$ each. 
The current analysis assumes the $\Upsilon$(1S) produced unpolarized, in agreement 
with previous CDF results (polarization parameter $\alpha$=-0.12$\pm$0.22 for 8$<p^{\Upsilon}_{T}<$20 GeV/$c$); 
thus, contribution to the systematic uncertainties is not included. 
Differential cross sections, normalized to unity, for the three rapidity ranges, 
are shown in figure ~\ref{fig:upsi_cross}: they are compared to theoretical predictions, 
that combined separate perturbative approacches for the low and high p$^{}_{T}$ regions.           
\section{Conclusions}
Studies of beauty production are important tests for perturbative QCD: results from CDF and D\O \, 
benefit from the very high statistic provided by the Tevatron collider and they 
are generally compatible with the recent theoretical predictions. Further improvements are expected in the near future and 
new measurements on $b\bar{b}$ cross section and correlations are already in progress. 



\section*{References}
\begin{thebibliography}{99}
\bibitem{mangano} S.Frixione, M.Mangano, Nucl. Phys.{\bf{ B 483}} (1997) 321, hep-ph/9605270.
\bibitem{mary} The CDF Collaboration, hep-ex/0412071, Published in Phys.Rev.D71:032001,2005.
\bibitem{me} CDF Public results in http://www-cdf.fnal.gov/physics/new/qcd/qcd$\_$plots/ \\
   bjets$\_$run2/blessed$\_$results.html
\bibitem{ups} The D\O \, Collaboration, hep-ex/0502030, FERMILAB-PUB-05/020-E; Submitted to PRL.
\end{thebibliography}
\end{document}